\documentclass{caosp309}

\usepackage{graphicx}

\usepackage{natbib}
\articleNo{B07}
\pubyear{2024}
\volume{54}
\volnumber{2}
\firstpage{70}
\received{November 1, 2023}
\accepted{November 28, 2023}

\begin{document}

\hauthor{C.I.\,Eze and G.\,Handler}

\title{Photometric sample of early B-type pulsators in eclipsing binaries observed with {\it TESS}}

\author{
        C.I.\,Eze\inst{1,2}\orcid{0000-0003-3119-0399}
      \and
        G.\,Handler\inst{1}\orcid{0000-0001-7756-1568}
       }

\institute{
           Nicolaus Copernicus Astronomical Center of the Polish Academy of Sciences, Warsaw, Poland \email{cheze@camk.edu.pl}
         \and 
           Department of Physics and Astronomy, University of Nigeria, Nsukka, Nigeria 
          }

\date{November 1, 2023}

\maketitle

\begin{abstract}
Asteroseismology coupled with eclipsing binary modelling shows a great potential in improving the efficiency of measurements or calibrations of the interior mixing profile in massive stars. This helps, for instance in treating the challenging and mysterious discrepancies between observations and models of  its stellar structure and evolution. This paper discusses the findings in our work titled {\it $\beta$ Cephei pulsators in eclipsing binaries observed with {\it TESS}}, which aimed to compile a comprehensive catalogue of $\beta$ Cep pulsators in eclipsing binaries. Seventy eight (78) pulsators of the $\beta$ Cep type in eclipsing binaries among which are 59 new discoveries were reported. Here, we also report a fresh  analysis of eight additional stars which were outside the scope of the earlier mentioned work. Six $\beta$ Cep pulsators in eclipsing binaries are reported, among which 5 are new discoveries and 1 is a confirmation of a candidate earlier suggested in literature. Our sample allows for future ensemble asteroseismic modelling of massive pulsators in eclipsing binaries to treat the discrepancy between observations and models. 
\keywords{asteroseismology -binaries: general - stars: evolution - massive stars: oscillations (including pulsations) - stars: rotation}
\end{abstract}

\section{Introduction}
\label{intr}

Asteroseismology is the study of the interior structure of stars via their pulsations \citep[e.g.,][]{Gough1985}. Massive stars also show pulsations and offer unique opportunities to constrain their properties via asteroseismology \citep{Aertsetal2010}. Three classes of massive pulsators exist, among which are $\beta$ Cephei stars.  Beta Cephei stars are main sequence stars with masses of approximately 9 -- 17 M$_{\sun}$ \citep[e.g.,][]{StankovandHandler2005}. They have low radial order pressure (p), gravity (g) and mixed modes, pulsation amplitudes up to a few tenths of magnitudes and pulsation periods of several hours (approximately 2 to 6 hours) \citep[e.g.,][]{StankovandHandler2005}. Studying massive pulsators in eclipsing binaries proves to be very beneficial as it combines the strengths of asteroseismology and eclipsing binary modelling to probe the stars.  Unfortunately, the number of such reported systems is much smaller compared to that of their low mass counterparts \citep{Kirketal2016, Pedersenetal2019}. This, however, impedes a more holistic asteroseismic study of massive stars.   

The present paper discusses the findings in the work titled {\it $\beta$ Cephei pulsators in eclipsing binaries observed with {\it TESS}} by Eze \& Handler (ApJS, submitted), which aimed to compile a comprehensive catalogue of early B-type (B0 -- B3) pulstators in eclipsing binaries observed by {\it TESS} with a particular focus on {\ensuremath{\beta}}~Cep pulsators, to probe the evolution and properties of massive stars by harnessing the combined potentials of eclipsing binary stars and asteroseismology. In this paper, we also conduct a fresh  analysis of eight additional stars which were outside the scope of our earlier mentioned work. Here, we briefly describe the analysis in Sect. 2, discuss the results in Sect. 3 and conclude in Sect. 4.

\section{Analysis}

 A total of 8055 stars of spectral types B0 -- B3 were analyzed in our referenced paper (see the paper for the details of the analysis and results).  Here, as already pointed out, we also analyzed eight additional stars of spectral types O8 -- B0 and B3 -- B5. The analysis was done  using the methods described in details in the referenced work.   The result is shown in Table \ref{t1}.

\begin{figure}
\centerline{\includegraphics[width=1.0\textwidth,clip=]{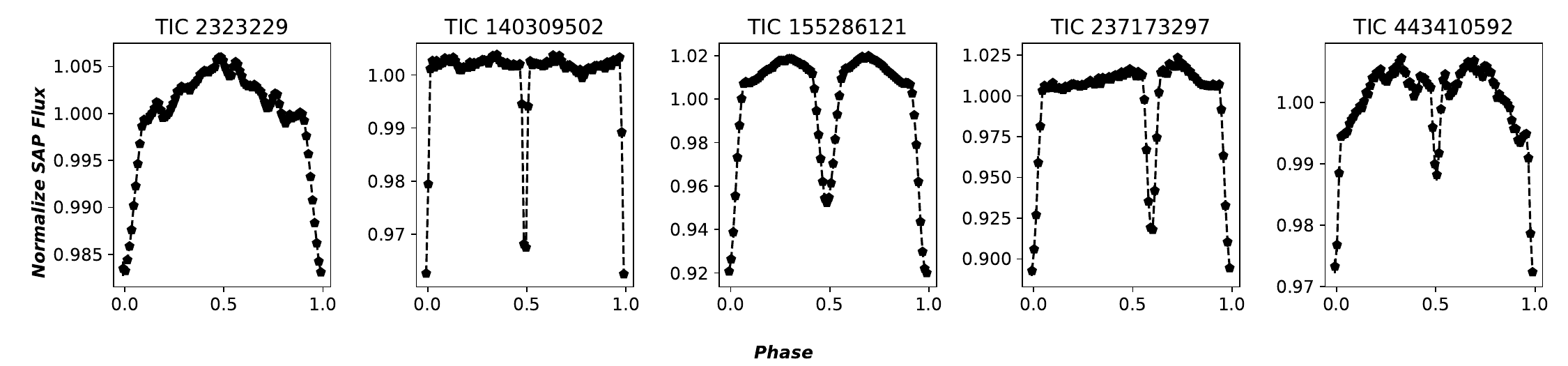}}
\caption{Binned Phase diagrams of light curves of the additional new definite and candidate $\beta$ Cep pulsators in eclipsing binaries after prewhitening the strongest pulsations. TIC 336660284 is not included in Figure 1 as it has insufficient data (an incomplete orbital cycle) to be accurately phased but showed first primary eclipse at 1606.35(30) BTJD. TIC 2323229, TIC 140309502 and TIC 443410592 are definite $\beta$ Cep pulsators in eclipsing binaries  whereas TIC 155286121, TIC 237173297 and TIC 336660284 are candidate $\beta$ Cep pulsators in eclipsing binaries. }
\label{f1}
\end{figure}

\begin{figure}
\centerline{\includegraphics[width=1.0\textwidth,clip=]{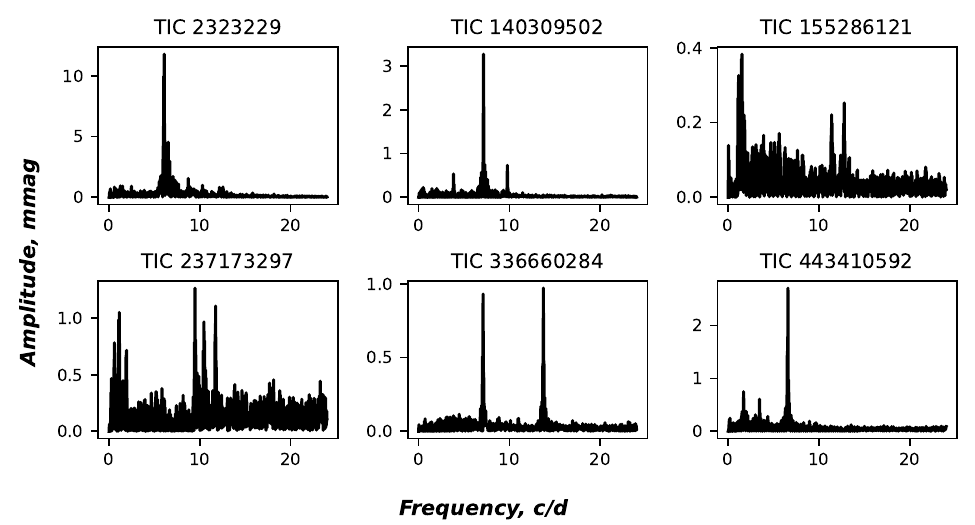}}
\caption{Fourier spectra of the additional new definite and candidate $\beta$ Cep pulsators in eclipsing binaries after removing the orbital light variations. Whereas TIC 2323229,          TIC 140309502 and TIC 443410592 are definite $\beta$ Cep pulsators in eclipsing binaries,  TIC 155286121, TIC 237173297 and TIC 336660284 are candidates.}
\label{f2}
\end{figure}

 We examined the {\it TESS} light curves \citep{Rickeretal2015} of the targets for independent pulsations via successive prewhitening of their Fourier spectra using the {\it Period04} software \citep{LenzandBreger2005} and accounted for blends in their {\it TESS} pixels using  Eleanor \citep{Feinsteinetal2019} and {\it TESS}-Localize \citep{HigginsandBell2022}.  In each case, we removed the effect of binarity by subtracting  the harmonic binary model from the light curves before the  pulsation analysis. The orbital periods of the stars with their analytical uncertainties are also determined using the {\it Period04} program \citep{LenzandBreger2005}. The binned phase diagrams and Fourier spectra of the stars' {\it TESS} light curves are shown in Figures \ref{f1} and \ref{f2} respectively.

\section{Results and discussion}

Eze \& Handler (ApJS, submitted) reported 78 pulsators of the $\beta$ Cephei type in eclipsing binaries, 59 of which are new discoveries. Ten ellipsoidal variables with $\beta$ Cep pulsating components are also reported in the paper. Here, we report three definite and three candidate $\beta$ Cep pulsators in eclipsing binaries as shown in Table \ref{t1}. The stars adjudged candidates have nearby contaminators and  signals that are too weak to be localized. They are denoted with $'?'$ in their variability in Table \ref{t1}. The binarity and  $\beta$ Cep variability  are reported for the first time here for the stars TIC 2323229 (HD 330666; Sp. type: B8), TIC 155286121 (HD 319702; Sp. type: O8III), TIC 237173297 (HD 329034; Sp. type: B3V) and TIC 336660284 (HD 125206; Sp. type: O9.7IVn). TIC 237173297 has an eccentric orbit and shows a likely reflection effect in its light curves.  The binarity of TIC 443410592 (HD 115282; Sp. type: B2III)  was first identified by \citet{Ijspeertetal2021},  but the $\beta$ Cep pulsations are reported here for the first time. \citet{Labadie-Bartzetal2020} first reported the $\beta$ variability of TIC 140309502 (CD-44 4484; Sp. type: B5) and also observed shallow eclipses in its light curves, for which they recommended further photometric confirmation. Using {\it TESS} photometry, this work confirms TIC 140309502 to be a definite  $\beta$ Cep pulsator in an eclipsing binary as both the eclipses and the $\beta$ Cep pulsations are localized to the position of this target. However, contrary to the orbital period of 13.721(7) d reported by \citet{Labadie-Bartzetal2020}, an orbital period of 4.649(5) or twice as much is obtained with the {\it TESS} photometry.  Two stars TIC 72211082 and TIC 141903541 with no eclipses are rejected here. They, however, show $\beta$ Cep pulsations with dominant pulsational frequencies 12.48585 c/d and 7.14297 c/d  respectively. TIC 72211082 appears more like a rotational or ellipsoidal variable with pulsating component(s).

\begin{table}[t]
\small
\begin{center}
\caption{The lists of the additional new definite and candidate $\beta$ Cep pulsators in eclipsing binaries not reported in the earlier work. $F_{d}$ is the dominant $\beta$ Cep pulsational frequency in c/d, A is the amplitude of the dominant pulsational frequency in mmag and S/N is the signal-to-noise ratio of the dominant  frequency  }
\label{t1}
\begin{tabular}{lllllll}
\hline\hline
TIC ID	&	Variability 	&	P (d)	&$F_{d}$ (c/d)	&	A (mmag) &	S/N	  \\
\hline
TIC 2323229     &	EB+bCep	&	1.0170(1)	&	6.0782 &	13.55&		81.5 \\
TIC 140309502   &	EB+bCep	&	9.298(5)	&		7.1382 &	3.51 &		78.6  \\
TIC 155286121 	&	EB+bCep+SPB?	&	2.01443(9)		&		11.4033 &		0.34&	 8.6\\
TIC 237173297 	&	EB+bCep+SPB?	&	2.9907(2)		& 11.7226 &		1.4 &	10.2  \\
TIC 336660284   &	EB+bCep?	&	13.35781(3)	&	13.7379 &	0.98 &		48.6 \\
TIC 443410592  	&	EB+bCep	&	4.2311(3)		&		6.5731 &		2.99 &		59.1  \\
\hline\hline
\end{tabular}
\end{center}
\end{table}

\section{ Conclusion}
In this paper, we report the findings from our work which analyzed 8055 stars for $\beta$ Cep pulsations in eclipsing binaries and that resulted in 78 pulsators of the $\beta$ Cephei type in eclipsing binaries, among which 59 are new discoveries. Here, six $\beta$ Cep pulsators in eclipsing binaries are also reported, five of which are new additional discoveries and one is a confirmation of a candidate earlier mentioned in literature. The present paper adds a few stars to the sample of {\ensuremath{\beta} Cep pulsators, reported until date, for a more general and homogeneous in depth asteroseismic analysis of massive stars. We have already started follow up of some of the most interesting candidates. The spectroscopic observations of some of the candidates have been conducted using four different instruments among which is Skalnate Pleso Observatory (SPO). Four $\beta$ Cep pulsators in eclipsing binaries listed in the referenced work were observed using the echelle spectrograph at SPO and are currently  under analysis.

\acknowledgements
This work was supported by the Polish National Science Foundation (NCN) under grant nr. 2021/43/B/ST9/02972 and used the {\it TESS} data obtained from the Mikulski Archive for Space Telescopes (MAST). This paper also made use of the SIMBAD data base and the VizieR catalogue access tool operated at CDS, Strasbourg, France;the SAO/NASA Astrophysics Data System. It also made use of data from the European Space Agency (ESA) mission
{\it Gaia}\footnote{https://www.cosmos.esa.int/gaia}, processed by the {\it Gaia}
Data Processing and Analysis Consortium (DPAC\footnote{https://www.cosmos.esa.int/web/gaia/dpac/consortium}). Funding for the DPAC
has been provided by national institutions, in particular the institutions
participating in the {\it Gaia} Multilateral Agreement.

\bibliography{caosp309}

\end{document}